\documentclass[12pt]{iopart}
\usepackage{graphicx}
\usepackage{color}
\expandafter\let\csname equation*\endcsname\relax

\expandafter\let\csname endequation*\endcsname\relax

\usepackage{amsmath}
\usepackage[caption=false]{subfig}
\usepackage[final]{changes}

\begin{document}

\title[]{A theoretical framework for comparing noise characteristics of spectral, differential phase-contrast and spectral differential phase-contrast X-ray imaging}

\author{Korbinian Mechlem, Thorsten Sellerer, Manuel Viermetz, Julia Herzen, Franz Pfeiffer}

\address{}
\ead{korbinian.mechlem@ph.tum.de}
\vspace{10pt}
\begin{indented}
\item[]January 2020
\end{indented}

\begin{abstract}
Spectral and grating-based differential phase-contrast X-ray imaging are two emerging technologies that offer additional information compared with conventional attenuation-based X-ray imaging. In the case of spectral imaging, energy-resolved measurements allow the generation of material-specific images by exploiting differences in the energy-dependent attenuation. Differential phase-contrast imaging uses the phase shift that an X-ray wave exhibits when traversing an object as contrast generation mechanism. Recently, we have investigated the combination of these two imaging techniques (spectral differential phase-contrast imaging) and demonstrated potential advantages compared with spectral imaging. In this work, we present a noise analysis framework that allows the prediction of (co-) variances and noise power spectra for all three imaging methods. Moreover, the optimum acquisition parameters for a particular imaging task can be determined. We use this framework for a performance comparison of all three imaging methods. The comparison is focused on (projected) electron density images since they can be calculated with all three imaging methods. Our study shows that spectral differential phase-contrast imaging enables the calculation of electron density images with strongly reduced noise levels compared with the other two imaging methods for a large range of clinically relevant pixel sizes. In contrast to conventional differential phase-contrast imaging, there are no long-range noise correlations for spectral differential phase-contrast imaging. This means that excessive low frequency noise can be avoided. We confirm the analytical predictions by numerical simulations.
\end{abstract}

%\maketitle

\section{Introduction}
Spectral and grating-based differential phase-contrast X-ray imaging are two emerging technologies that offer additional information compared to conventional attenuation-based X-ray imaging.\\
In the case of spectral imaging, information about the energy-dependent attenuation of an object is obtained by acquiring measurements with two or more photon energy spectra. Material decomposition algorithms use the energy-resolved measurements to generate basis material images, which can provide information about the chemical composition of an object. Various dual-energy acquisition schemes and  recent advances in photon-counting detector technology \cite{Ballabriga2016,willemink2018photon} have encouraged research towards medical applications of spectral X-ray imaging \cite{mccollough2015dual,nadjiri2018spectral,symons2018photon,schwaiger2018three}. \added{The ability of photon-counting detectors to acquire multiple energy-resolved measurements that are spatially and temporally registered has proven advantageous for multi-material decomposition \cite{mechlem2018spectral,symons2017photon,dangelmaier2018experimental}.}\\
Grating-based differential phase-contrast (DPC) imaging exploits an entirely different contrast generating mechanisms in addition to the conventional attenuation contrast: The phase-contrast image is obtained by indirectly measuring the (differential) phase shift that an X-ray wave exhibits when transversing an object \cite{pfeiffer2006phase,pfeiffer2007hard}. This differential phase shift can be directly related to the projected electron density (PED) of the object. The dark-field image is generated by ultra-small angle X-ray scattering from microstructures \cite{pfeiffer2008hard}. \added{It is connected to the real space autocorrelation function \cite{yashiro2010origin,strobl2014general} and thus provides information about the internal structure or surface properties \cite{yashiro2019probing} of an object far below the detector resolution.} Laboratory experiments have demonstrated that the phase-contrast image can provide a strongly improved contrast-to-noise ratio compared with attenuation-based imaging \cite{herzen2009quantitative,zambelli2010radiation,donath2010toward}. However, based on a theoretical noise analysis, doubts have been raised whether these improvements can be transferred  to clinical computed tomography (CT) applications \cite{raupach2011analytical,raupach2012performance}. Experimental studies have demonstrated that similarly to spectral imaging, material-specific information can be extracted from phase-contrast measurements \cite{braig2018direct}. \\
Recently, we have investigated the combination of grating-based DPC radiography and spectral radiography \cite{mechlem2019spectral} and have developed a basis material decomposition algorithm that uses the spectral and the phase contrast information simultaneously. Our numerical experiments have demonstrated that spectral differential phase-contrast (SDPC) radiography yields quantitatively correct basis material images with strongly reduced noise levels compared with conventional spectral imaging.\\
Taking the characteristics of all three aforementioned imaging methods into account, the question of the optimal method for a particular imaging task arises. In this work, we investigate one important aspect of this question by conducting an in-depth  noise analysis of all three imaging methods. Towards this end, we develop a noise analysis framework based on the Cram\'{e}r Rao lower bound (CRLB) \cite{kay1993fundamentals}. This framework allows the determination of the optimum imaging parameters for all three methods and the prediction of noise correlations and noise power spectra. Since all three aforementioned imaging methods can determine the PED, we focus on PED images for our comparison. Although only projection imaging is considered in this work, a generalization to 3D computed tomography is possible.  \added{Using an imaging task that can be viewed as a simplified version of human X-ray radiography, we demonstrate that the combination of spectral and phase-contrast imaging has the potential to generate PED images with strongly reduced noise levels compared with the individual methods.} Our theoretical analysis suggests that SDPC imaging outperforms spectral and DPC imaging for a large range of clinically relevant pixel sizes.  Finally, we discuss the noise characteristics of SDPC imaging in the context of the comparison between attenuation-based imaging and DPC imaging by Raupach and Flohr \cite{raupach2011analytical}. 

\section{Methods}
Before explaining the derivation of the noise analysis framework based on the CRLB, we give a brief overview of the physical models (forward models) of the measurement acquisition processes for all three imaging methods under consideration (spectral, SDPC and DPC imaging). Furthermore, we formulate the signal extraction process as a maximum-likelihood (ML) estimation problem for all three imaging methods. ML estimation has already been used successfully for projection-based material decomposition in spectral imaging \cite{schlomka2008experimental,ehn2017basis}. In this case, the ML estimator has many desirable properties such as the capability to handle overdetermined systems (i.e. having more energy bins than basis materials). Moreover, it is unbiased and efficient (i.e. it achieves the minimum variance for an unbiased estimator) in the limit of low noise levels. We also successfully applied an ML-based decomposition algorithm to SDPC imaging \cite{mechlem2019spectral}. In the case of DPC imaging, Fourier processing \cite{pfeiffer2008hard} is the most commonly used method of signal extraction due to the comparatively low computational complexity. \added{Using error propagation, the noise properties of DPC imaging in combination with Fourier processing have already been investigated for CT reconstruction \cite{chen2011scaling} and in the projection domain \cite{chabior2011signal,revol2010noise,yashiro2008efficiency}.} However, it has been demonstrated that  a Fourier-based estimator for DPC imaging is not efficient \cite{ge2014cramer}. For this reason, and to make the signal extraction for all three methods more comparable, we use an ML estimator for DPC imaging. This estimator is  similar to weighted least-squares signal extraction methods for DPC imaging \cite{marschner2016helical,kaeppler2017improved}. \\
In the following, we assume that a photon-counting detector (PCD) is used to acquire energy-resolved measurements for all three imaging methods. The number of registered photon counts is therefore modeled by a Poisson distribution. Moreover, we assume that there is no correlation between the photon counts registered in different detector pixels or energy bins. These assumptions are valid for an ideal PCD, but current real PCDs exhibit various undesired sensor effects (e.g. pulse pile-up, charge sharing) that could cause a violation of these assumptions \cite{rajbhandary2017effect,wang2011pulse}. In the limit of negligible electronic noise and detector blur, the noise analysis presented in this work could be extended to other spectral imaging methods, such as dual source CT or kVp-switching. As will be explained later, the differential phase shift for SDPC and DPC imaging couples pixels in the direction perpendicular to the orientation of the grating bars. However, the pixels remain uncoupled in the direction parallel to the grating bars. For simplicity, we will therefore consider a one-dimensional PCD in the following because each detector row can be treated separately for all three imaging methods.

\subsection{Forward models and signal extraction}
Basis material decomposition for spectral imaging relies on the assumption that the energy-dependent attenuation of any material can be modeled by a linear combination of a few basis materials.
Neglecting materials with K-edges in the relevant energy range for medical imaging ($\approx 20-140 \ \rm{keV}$), only two basis materials are needed. This is a consequence of the fact that there are only three interaction mechanisms (photoelectric effect, Compton-scattering and Rayleigh-scattering) in this energy range. Furthermore, the contribution of Rayleigh-scattering to the total attenuation cross-section is typically small compared with the other two interaction mechanisms. For simplicity, we will focus on two basis materials in the following.
For spectral imaging, the expected number of photon counts $\hat{y}_i^s$ registered in energy bin $s$ and detector pixel $i$ is thus modeled by \cite{schlomka2008experimental}:
\begin{equation}
\hat{y}_i^s = \int_0^{\infty} t(\mathcal{E}) \mathcal{R}^s(\mathcal{E}) e^{-A_1^i f_1(\mathcal{E}) - A_2^i f_2(\mathcal{E})} d\mathcal{E},    
\end{equation}
where $t(\mathcal{E})$ is the source spectrum and $\mathcal{R}^s(\mathcal{E})$ describes the detector response, i.e. the probability that a photon with energy $\mathcal{E}$ is detected by energy bin $s$ of the PCD. The functions $f_1(\mathcal{E})$ and $f_2(\mathcal{E})$ represent the energy-dependent attenuation of the two basis materials and the corresponding basis material line integrals (for detector pixel $i$) are denoted by $A_1^i$ and $A_2^i$. \\
Assuming uncorrelated Poisson statistics, the negative log-likelihood function for spectral imaging is given by:
\begin{equation}
\label{ml_spectral}
-L(\vec{A}_1, \vec{A}_2) = \sum_{i=1}^N \sum_{s=1}^S \hat{y}_i^s - y_i^s {\rm{ln}} \left( \hat{y}_i^s \right),
\end{equation}
where $N$ and $S$ represent the number of detector pixels and energy bins of the PCD, respectively. The quantity $y_i^s$ denotes the measured number of photon counts. ML decomposition
is performed by finding the basis material line integrals $\vec{A}_1  = \left(A_1^1,...,A_1^N \right), \ \vec{A}_2  = \left(A_2^1,...,A_2^N \right)$ that minimize $-L(\vec{A}_1, \vec{A}_2)$. This optimization problem can be solved separately for each detector pixel. The PED ($\rho_{\rm{e}}$) can be calculated by a linear combination of the basis material line integrals:
\begin{equation}
\label{ped_basis_mat_approx}
\rho_{\rm{e}}^i =  A_1^i \rho_{\rm{e}}^{\rm{V}}(\rm{M}_1) + A_2^i \rho_{\rm{e}}^{\rm{V}}(\rm{M}_2),
\end{equation}
where $\rho_{\rm{e}}^{\rm{V}}(\rm{M}_1)$ and $\rho_{\rm{e}}^{\rm{V}}(\rm{M}_2)$ are the volume electron densities of the two basis materials. \added{The idea of approximating the electron density of any material by a linear combination of the two basis materials is conceptually very similar to the standard dual energy assumption of modeling the energy-dependent attenuation with two basis materials. Consequently, equation \ref{ped_basis_mat_approx} is also only valid for the dual energy parameter range (low-Z materials, energy range $\approx 20 - 140 \ \rm{keV}$).} 
For DPC imaging, the three contrast modalities (attenuation, differential phase shift and dark-field) are extracted from stepping curve measurements that are generated by shifting one of the gratings \cite{pfeiffer2006phase, pfeiffer2008hard}. In contrast to spectral and SDPC imaging, no energy-resolved measurements are acquired. We therefore assume a PCD with just one threshold for DPC imaging. The stepping curve is typically modeled by a cosine (or sine) function \cite{pfeiffer2008hard} and beam hardening effects caused by the polychromatic spectrum are neglected.  The expected intensity $\hat{y}_i^r$ for stepping position $r$ and detector pixel $i$ can thus be written as: 
\cite{brendel2016penalized}:
\begin{equation}
\label{conv_dpc_fwd}
\hat{y}_i^r = b e^{-\mu_i} \left(1 + V e^{-\epsilon_i} \cos \left( \phi_r + \Delta \phi_i \right) \right),
\end{equation}
where  $\phi_r$ and $V$ are the reference phase (for step $r$) and the reference visibility of the stepping curve, respectively. The quantities $\mu_i$, $\Delta \phi_i$ and $\epsilon_i$ describe the attenuation of the object, the phase shift of the stepping curve and the visibility reduction (dark-field signal), respectively. The reference intensity measured with an open beam is given by the parameter $b$.
\added{The standard DPC stepping curve model (equation \ref{conv_dpc_fwd}) implicitly represents the polychromatic spectrum by an effective X-ray energy. This approximation is reasonably accurate for weakly attenuating samples that only slightly distort the incident X-ray spectrum. Since enhancing soft tissue contrast is one of the main application cases for DPC imaging, this assumption is often justified.} 
The attenuation, differential phase and dark-field images ($\vec{\mu}, \Delta \vec{\phi}, \vec{\epsilon}$) are calculated by minimizing the negative log-likelihood of the measured data:
\begin{equation}
\label{ml_dpc}
-L \left(\vec{\mu}, \Delta \vec{\phi}, \vec{\epsilon} \right) = \sum_{i=1}^N \sum_{r=1}^R \hat{y}_i^r - y_i^r {\rm{ln}}\left(\hat{y}_i^r   \right),
\end{equation}
where $R$ is the number of stepping positions and $y_i^r$ denotes the number of photon counts measured for detector pixel $i$ and stepping position $r$. Similarly to spectral imaging, this optimization problem is separable with respect to the detector pixels.  From the differential phase shift, the gradient of the PED can be calculated:
\begin{equation}
\label{delta_phi_from_rhoe_dpc}
\Delta \phi_i = \mathcal{S} \left(\rho_{\rm{e}}^{i+1} - \rho_{\rm{e}}^{i}\right) ,
\end{equation}
where the sensitivity $\mathcal{S}$ represents the conversion factor between a PED difference and the corresponding phase shift of the stepping curve. The higher the sensitivity, the larger the phase shift for a given electron density difference between two neighboring pixels.
Assuming that the sample is placed between the G1 and G2 grating (compare figure \ref{fig:setup_geometry}), the sensitivity is given by \cite{donath2009inverse}:
\begin{equation}
\label{sens_dpc_def}
\mathcal{S} =  \frac{r_e d}{p_2 a} \left( 1 - \frac{l}{d} \right) {\left(\frac{hc}{\mathcal{E}_{\rm{eff}}} \right)}^2 ,
\end{equation}
where $r_{\rm{e}}$ is the classical electron radius, $d$ is the distance between the G1 and G2 grating, $l$ is the distance between the G1 grating and the object and $p_2$ is the period of the G2 grating. The parameter $a$ represents the effective pixel size (i.e. the detector pixel size divided by the geometrical magnification of the setup) and $\mathcal{E}_{\rm{eff}}$ is the effective energy of the setup. In analogy to attenuation-based imaging, it can be defined as \cite{chabior2011beam}:
\begin{equation}
\label{E_eff_dpc}
\mathcal{E}_{\rm{eff}} = {\left( \frac{\int_0^\infty t(\mathcal{E})\mathcal{R}(\mathcal{E}) e^{-\mu(\mathcal{E})}  V(\mathcal{E})  \mathcal{E}^{-2} d\mathcal{E}}{\int_0^\infty t(\mathcal{E})\mathcal{R}(\mathcal{E}) e^{-\mu(\mathcal{E})} V(\mathcal{E})  d\mathcal{E}} \right)}^{-\frac{1}{2}},
\end{equation}
where $\mu(\mathcal{E})$ and $V(\mathcal{E})$ represent the energy-dependent attenuation of the object and the energy-dependent visibility of the interferometer, respectively. Assuming  a known value of the PED at the left and right edges of the detector (for simplicity we assume that the open beam hits the edges of the detector, i.e. $\rho_{\rm{e}}^1 = \rho_{\rm{e}}^N = 0$), the PED for an arbitrary pixel can be calculated by integration of the differential phase shifts:
\begin{equation}
\label{dpc_int_l_to_r}
\rho_{\rm{e}}^i = \frac{1}{\mathcal{S}} \sum_{q=1}^{i-1} \Delta \phi_q.
\end{equation}
Alternatively, it would be possible to integrate starting from the right side of the detector:
\begin{equation}
\label{dpc_int_r_to_l}
\rho_{\rm{e}}^i = - \frac{1}{\mathcal{S}} \sum_{q=i}^{N-1} \Delta \phi_q.
\end{equation}
However, these integration strategies are suboptimal for our goal of providing an in-depth noise analysis of all three imaging methods. The summation introduces a strong spatial dependency of the PED variance, even for a homogeneous sample \added{(i.e. a sample that produces the same expected stepping curve measurements for each detector pixel)}. In this case, the variance of the PED increases linearly from left to right or right to left (if using equation (\ref{dpc_int_l_to_r}) or equation (\ref{dpc_int_r_to_l}), respectively). However, error propagation calculations show (see appendix) that the variance remains constant (even for nonhomogeneous samples) if the electron density is calculated as the average of both summation strategies:
\begin{equation}
\label{dpc_int_lr}
\rho_{\rm{e}}^i = \frac{1}{2\mathcal{S}} \left( \sum_{q=1}^{i-1} \Delta \phi_q-  \sum_{q=i}^{N-1} \Delta \phi_q \right).
\end{equation}
In the appendix, we show that the electron density variance is reduced by a factor of two when compared to variance obtained using equation (\ref{dpc_int_l_to_r}) or (\ref{dpc_int_r_to_l}). \\
SDPC imaging can be viewed as a combination of spectral and DPC imaging. The spectral and phase contrast information is used simultaneously by acquiring energy-resolved stepping curve measurements. The expected number of photon counts $\hat{y}_i^{rs}$ for detector pixel $i$, stepping position $r$ and energy bin $s$ can be modeled as:
\begin{equation}
\label{sdpc_model}
\begin{split}
&\hat{y}_i^{rs} = \int_0^\infty t(\mathcal{E})\mathcal{R}^s(\mathcal{E}) e^{- A_1^i f_1(\mathcal{E}) - A_2^i f_2(\mathcal{E})}\\
&\left[1 + V(\mathcal{E}) e^{-d_{\epsilon}^i f_{\epsilon}(\mathcal{E})} \cos \left(\phi_r(\mathcal{E}) + \Delta \phi_i (\mathcal{E}) \right) \right] d\mathcal{E},
\end{split}
\end{equation}
where $d_{\epsilon}^i$ is the line integral of an artificial dark-field basis material (see \cite{mechlem2019spectral} for a more detailed explanation) and $f_{\epsilon}(\mathcal{E})$ is the corresponding energy-dependency of the dark-field signal. Compared to the forward model of DPC imaging (equation \ref{conv_dpc_fwd}), the polychromatic spectrum is taken into account and thus the visibility $V(\mathcal{E})$, reference phase $\phi_r(\mathcal{E})$ and the phase shift $\Delta \phi_i (\mathcal{E})$  become energy-dependent.
In a real experiment, all setup-dependent quantities ($t(\mathcal{E}), \mathcal{R}^s(\mathcal{E}), V(\mathcal{E}), \phi_r(\mathcal{E})$) can depend on the spatial position and therefore on the detector pixel index $i$. For simplicity and clarity, we have omitted this possible dependency in this section.
Similarly to DPC imaging, the phase shift depends on the gradient of the PED:
\begin{equation}
\label{delta_phi_from_rhoe_sdpc}
\Delta \phi_i (\mathcal{E}) =  \frac{r_{\rm{e}} d}{p_2 a}\left( 1 - \frac{l}{d} \right) {\left(\frac{hc}{\mathcal{E}} \right)}^2 \left({\rho}_{\rm{e}}^{i+1} - {\rho}_{\rm{e}}^{i}\right) \equiv \mathcal{S}(\mathcal{E}) \left({\rho}_{\rm{e}}^{i+1} - {\rho}_{\rm{e}}^{i}\right),
\end{equation}
where $\mathcal{S}(\mathcal{E})$ is the energy-dependent sensitivity of the setup.
The key idea behind connecting spectral and DPC imaging and eliminating the PED as an additional optimization variable  is expressing the PED as the sum of the projected basis material electron densities:\begin{equation}
\label{ed_modeling}
{\rho}_e^i = A_1^i  \rho_{\rm{e}}^{\rm{V}}(\rm{M}_1) + A_2^i  \rho_{\rm{e}}^{\rm{V}}(\rm{M}_2).
\end{equation}
Two basis materials images ($\vec{A}_1, \vec{A}_2$) and a dark-field ($\vec{d}_{\epsilon}$) image can be reconstructed by minimizing the following negative log-likelihood function:
\begin{equation}
\label{ml_sdpc}
-L(\vec{A}_1, \vec{A}_2, \vec{d}_{\epsilon}) = \sum_{i=1}^N \sum_{s=1}^S \sum_{r=1}^R \hat{y}_i^{rs} - y_i^{rs} \text{ln}\left( \hat{y}_i^{rs} \right).
\end{equation}
As the forward model depends on the basis material thicknesses and their spatial gradients (via the gradient of the PED), the log-likelihood cannot be optimized separately for each detector pixel.

\subsection{Noise analysis with the Cram\'{e}r Rao lower bound (CRLB)}
The CRLB is a powerful tool from estimation theory that predicts a lower bound for the variance of an unbiased estimator. Given a parameter vector $\vec{a}$, it can be shown that \cite{scharf1991statistical}:
\begin{equation}
\label{fisher_matrix_ineq}
\boldsymbol{C}(\vec{a}) - [\boldsymbol{F}(\vec{a})]^{-1} \geq 0,
\end{equation}
i.e. the matrix $\boldsymbol{C}(\vec{a}) - [\boldsymbol{F}(\vec{a})]^{-1}$ is positive semidefinite. Here, $\boldsymbol{C}(\vec{a})$ is the covariance matrix of $\vec{a}$:
\begin{equation}
\boldsymbol{C}_{uv} = {\rm{E}} \left[ (a_u - E(a_u)) (a_v - E(a_v)) \right],
\end{equation}
where $E(\cdot)$ denotes the expectation value. The Fisher information matrix $\boldsymbol{F}(\vec{a})$ is the expectation value of the curvature of the negative log-likelihood function:
\begin{equation}
{\boldsymbol{F}}_{uv} =  {\rm{E}} \left[- \frac{\partial^2 L(\vec{a})}{\partial a_u \partial a_v} \right].
\end{equation}
From equation \ref{fisher_matrix_ineq}, a lower bound for the variance of the estimated parameters can be deduced:
\begin{equation}
\label{var_lower_bound}
\boldsymbol{C}_{uu} = \sigma^2(a_u) \geq {(\boldsymbol{F}^{-1}})_{uu}.
\end{equation}
Since the ML estimator is unbiased and achieves the CRLB in the limit of low noise levels, the CRLB can be used to predict the noise levels for all three imaging methods under consideration. As will be shown later, the CRLB is a good predictor of the noise levels for clinically realistic photon statistics.\\ 
In the case of spectral imaging, the optimization problem is separable with respect to different pixels. We therefore obtain $N$ $2 \times 2$ (inverse) Fisher matrices. For simplicity, the dependency on the pixel index $i$ is suppressed in the following. As derived by Roessel and Herrmann \cite{roessl2009cramer}, the Fisher matrix is given by:
\begin{equation}
\boldsymbol{F}_{uv} = \sum_{s=1}^S \frac{1}{\hat{y}^s} \frac{\partial \hat{y}^s}{\partial A_u} \frac{\partial \hat{y}^s}{\partial A_v},
\end{equation}
where $u,v \in (1,2)$ are the basis material indices and
\begin{equation}
\frac{\partial \hat{y}^s}{\partial A_u} = - \int_0^{\infty} t(\mathcal{E}) \mathcal{R}^s(\mathcal{E}) e^{-A_1 f_1(\mathcal{E}) - A_2 f_2(\mathcal{E})} f_u(\mathcal{E}) d\mathcal{E}.   
\end{equation}
The elements of the Fisher matrix can be rewritten in a slightly more intuitive form:
\begin{equation}
\boldsymbol{F}_{uv} = \sum_{s=1}^S \hat{y}^s \bar{f}_u^s \bar{f}_v^s, \quad \bar{f}_u^s = \frac{1}{\hat{y}^s} \frac{\partial \hat{y}^s}{\partial A_u},
\end{equation}
where $\bar{f}_u^s$ is the weighted average attenuation caused by basis material $u$ in energy bin $s$. The weights for each energy bin are determined by the effective spectra (including object attenuation). The lower bound for the variances of the basis material line integrals (see equation \ref{var_lower_bound}) is then calculated using the analytical inversion formula for $2\times 2$ matrices. From the variances of the basis material thicknesses, the variance of the PED can be calculated by standard error propagation:
\begin{equation}
\sigma^2(\rho_{\rm{e}}) = \rho^{\rm{V}}_{\rm{e}}(\rm{M}_1)^2 \sigma^2(A_1) + \rho^{\rm{V}}_{\rm{e}}(\rm{M}_2)^2 \sigma^2(A_2) + 2\rho^{\rm{V}}_{\rm{e}}(M_1) \rho^{\rm{V}}_{\rm{e}}(M_2) {\rm{Cov}}(A_1, A_2),
\end{equation}
where the covariance of the two basis material thicknesses ${\rm{Cov}}(A_1,A_2)$ is estimated by $(\boldsymbol{F}^{-1})_{12}$. \\
Similarly to the derivation for spectral imaging \cite{roessl2009cramer}, The Fisher matrix for DPC imaging can be calculated as:
\begin{equation}
\boldsymbol{F}_{uv} (\vec{a}) = \sum_{r=1}^R \frac{1}{\hat{y}^r} \frac{\partial \hat{y}^r}{\partial a_u} \frac{\partial \hat{y}^r}{\partial a_v}, \quad \vec{a} = {(\mu, \Delta \phi, \epsilon)}^T.
\end{equation}
Explicitly calculating the gradients leads to:
\begin{equation}
\boldsymbol{F} =   \sum_{r=1}^R \frac{1}{\hat{y}^r}
\left( \begin{matrix}
\left(\hat{y}_r\right)^2 &  \hat{y}^r Q \sin(\phi^{\rm{eff}}_r) &   \hat{y}^r Q \cos(\phi^{\rm{eff}}_r )\\
\hat{y}^r Q \sin(\phi^{\rm{eff}}_r) &   Q^2 {\sin}^2(\phi^{\rm{eff}}_r) & Q^2 \sin(\phi^{\rm{eff}}_r) \cos(\phi^{\rm{eff}}_r) \\
  \hat{y}^r Q \cos(\phi^{\rm{eff}}_r ) & Q^2 \sin(\phi^{\rm{eff}}_r) \cos(\phi^{\rm{eff}}_r) & Q^2 {\cos}^2(\phi^{\rm{eff}}_r)
\end{matrix} \right),
\end{equation}
where $\phi^{\rm{eff}}_r = \phi_r - \Delta \phi$ is the effective phase and $Q= be^{-\mu}Ve^{-\epsilon}$ is the expected intensity ($be^{-\mu}$) multiplied by the effective visibility ($Ve^{-\epsilon}$)  of the stepping curve.
We are particularly interested in the lower bound for the variance of the differential phase shift $\sigma^2(\Delta \phi)$, as the PED is calculated by integration of the differential phase shifts (compare equation \ref{dpc_int_lr}). It is given by:
\begin{equation}
\sigma^2(\Delta \phi) \geq \left(\boldsymbol{F}^{-1}\right)_{22}. 
\end{equation}
For standard phase stepping with $R>3$ equidistantly distributed steps, the off-diagonal elements $\boldsymbol{F}_{12}$, $\boldsymbol{F}_{21}, \boldsymbol{F}_{13}, \boldsymbol{F}_{31}$ vanish \cite{weber2011noise}.
Moreover, numerical evaluations show that $\boldsymbol{F}_{23}$ and $\boldsymbol{F}_{32}$ are small compared with the corresponding diagonal entries and vanish in the limit of $R \rightarrow \infty$.
In this case, a simple interpretation of the lower bound for $\sigma^2(\Delta \phi)$ is possible:
\begin{equation}
\label{sigma_delta_phi_interp}
\sigma^2(\Delta \phi) \geq \frac{1}{\boldsymbol{F}_{22}} \propto \left[{Rbe^{-\mu}}{\left(Ve^{-\epsilon}\right)^2}\right]^{-1},
\end{equation}
i.e. $\sigma^2(\Delta \phi)$ is inversely proportional to the number of phase steps, the average number of photon counts per step and the effective visibility squared. The same result was obtained for a least-squares estimator (instead of an ML estimator) \cite{weber2011noise}. By applying standard error propagation techniques to equation \ref{dpc_int_lr}, the variance of the PED can be calculated from the variance of the differential phase shift:
\begin{equation}
\label{error_prop_delta_phi_rhoe}
\sigma^2 (\rho_{\rm{e}}^i) =   \frac{1}{4\mathcal{S}^2} \sum_{q=1}^{N-1} \sigma^2(\Delta \phi_q).
\end{equation}
For SDPC imaging, the optimization problem cannot be solved separately for each detector pixel. Consequently, there are 3N optimization variables, which we summarize in the vector $\vec{a} = {\left(A_1^1,...,A_1^N,A_2^1,...,A_2^N,d_{\epsilon}^1,...,d_{\epsilon}^N \right)}^T$
The Fisher matrix has $3N \times 3N$ entries that are calculated similarly to the other two imaging methods:
\begin{equation}
\boldsymbol{F}_{uv} = \sum_{i=1}^N \sum_{s=1}^S \sum_{r=1}^R \frac{1}{\hat{y}_i^{rs}} \frac{\partial \hat{y}_i^{rs}}{\partial a_u} \frac{\partial \hat{y}_i^{rs}}{\partial a_v}
\end{equation}
However, most of the elements of the Fisher matrix are zero because the differential phase shift only couples neighboring pixels. To write the partial derivatives of the forward model more compactly, we define the following abbreviations:
\begin{equation}
\begin{split}
&\alpha_{i}^{s}(\mathcal{E}) = t(\mathcal{E}) \mathcal{R}^s(\mathcal{E}) e^{-A_1^i f_1(\mathcal{E}) - A_2^i f_2(\mathcal{E})} \\
&\beta_i^{r}(\mathcal{E}) = V(\mathcal{E})e^{-d_{\epsilon}^i f_{\epsilon}(\mathcal{E})} \cos\left(\phi_r(\mathcal{E}) + \Delta \phi_i(\mathcal{E}) \right) \\
&\gamma_i^r(E) = V(\mathcal{E})e^{-d_{\epsilon}^i f_{\epsilon}(\mathcal{E})} \sin\left(\phi_r(\mathcal{E}) + \Delta \phi_i(\mathcal{E}) \right) 
%&\mathcal{S}(E) = \frac{r_e d}{p_2 a} {\left(\frac{hc}{\mathcal{E}} \right)}^2
\end{split}
\end{equation}
The non-zero partial derivatives of the forward model ($\hat{y}_i^{rs}$) are given by:
\begin{equation}
\begin{split}
\frac{\partial \hat{y}_i^{rs}}{\partial A_u^i} &= \int_0^\infty -\alpha_i^{s}(\mathcal{E}) \left(1 + \beta_i^r(\mathcal{E}) \right) f_u(\mathcal{E})  + \alpha_i^s(\mathcal{E}) \gamma_i^r(\mathcal{E}) \mathcal{S}(\mathcal{E}) \rho_{\rm{e}}(\rm{M}_u)d\mathcal{E}, \quad u \in (1,2)\\
\frac{\partial \hat{y}_i^{rs}}{\partial A_u^{(i+1)}} &= - \int_0^\infty \alpha_i^{s}(\mathcal{E})  \gamma_i^r(\mathcal{E}) \mathcal{S}(\mathcal{E}) \rho_{\rm{e}}(\rm{M}_u)d\mathcal{E}, \quad u \in (1,2) \\
\frac{\partial \hat{y}_i^{rs}}{\partial d_\epsilon^i} &=  -\int_0^\infty \alpha_i^{s}(\mathcal{E})  \beta_i^r(\mathcal{E}) f_\epsilon(\mathcal{E}) d\mathcal{E}. \quad
\end{split}
\end{equation}
The Fisher matrix for SDPC imaging must be inverted numerically to calculate the CRLB.
\subsection{Prediction of covariances and noise power spectra}
\label{sec:cov_and_nps_from_crlb}
In this section, we explain our approach for predicting covariances and noise power spectra for all three imaging methods.
Given an estimate of the covariances of the projected electron densities between different pixels, the noise power spectrum (NPS) can be estimated as:
\begin{equation}
\label{nps_from_cov}
{\rm{NPS}}(k) = \sum_{u=0}^{N-1} \sum_{v=0}^{N-1} e^{2\pi j \frac{v-u}{N}} {\rm{Cov}}( {\rho}_{\rm{e}}^{(u+1)}, {\rho}_{\rm{e}}^{(v+1)}),
\end{equation}
where $k$ is the spatial frequency and $j$ is the imaginary unit. In the appendix, we give a quick derivation of this result. For spectral imaging, the material decomposition is conducted separately for each pixel. This means that there are no noise correlations between different pixels:
\begin{equation}
{\rm{Cov}} \left(\rho_{\rm{e}}^{u}, \tilde{\rho}_{\rm{e}}^{v}\right) = 0, u \neq v , \quad {\rm{Cov}} \left(\rho_{\rm{e}}^{u}, \rho_{\rm{e}}^{u} \right) = \sigma^2(\rho_{\rm{e}}^u).
\end{equation}
For DPC imaging, the differential phase shifts are also uncorrelated. The covariance matrix of the differential phase shifts $\boldsymbol{C}^{\Delta \phi}$ is thus a diagonal matrix with diagonal elements $\boldsymbol{C}^{\Delta \phi}_{ii} = \sigma^2(\Delta \phi_i)$.  However, the integration step that is necessary to obtain the projected electron densities (compare equation \ref{dpc_int_lr}) introduces noise correlations. Using error propagation, the covariance matrix $C^{\rho_{\rm{e}}}$ for the projected electron densities can be calculated from the covariance of the differential phase shifts:
\begin{equation}
\boldsymbol{C}^{\rho_{\rm{e}}} = \boldsymbol{B} \boldsymbol{C}^{\Delta \phi} \boldsymbol{B}^T,
\end{equation}
where $\boldsymbol{B}$ is the transformation matrix between phase shifts ($\Delta \vec{\phi}$) and projected electron densities ($\vec{\rho}_e$):
\begin{equation}
\vec{\rho}_e = \boldsymbol{B} \vec{\phi}.
\end{equation}
The entries of the matrix $\boldsymbol{B}$ can be deduced form equation \ref{dpc_int_lr}. 
In the case of SDPC imaging, we use the entries of the inverse Fisher matrix as an estimate for the covariances:
\begin{equation}
{\rm{Cov}}(a^u, a^v) \approx \boldsymbol{F}^{-1}_{uv}, \quad \vec{a} = {\left(A_1^1,...,A_1^N,A_2^1,...,A_2^N,d_{\epsilon}^1,...,d_{\epsilon}^N \right)}^T. 
\end{equation}
Similarly to DPC imaging, the covariance matrix for the projected electron densities is calculated as:
\begin{equation}
\boldsymbol{C}^{\rho_{\rm{e}}} = \tilde{\boldsymbol{B}} \boldsymbol{F}^{-1} \tilde{\boldsymbol{B}}^T, \quad \vec{\rho}_e = \tilde{\boldsymbol{B}} \vec{a}.
\end{equation}
The entries of the transformation matrix $\tilde{\boldsymbol{B}}$ can be deduced from equation \ref{ed_modeling}.

\subsection{Numerical simulation}
\begin{figure}[h]
    \centering
    %\captionsetup[subfloat]{farskip=0pt,captionskip=0pt}
   \includegraphics[trim={0.0cm 0.0cm 0.0cm 0.cm},clip,width = 5.0in]{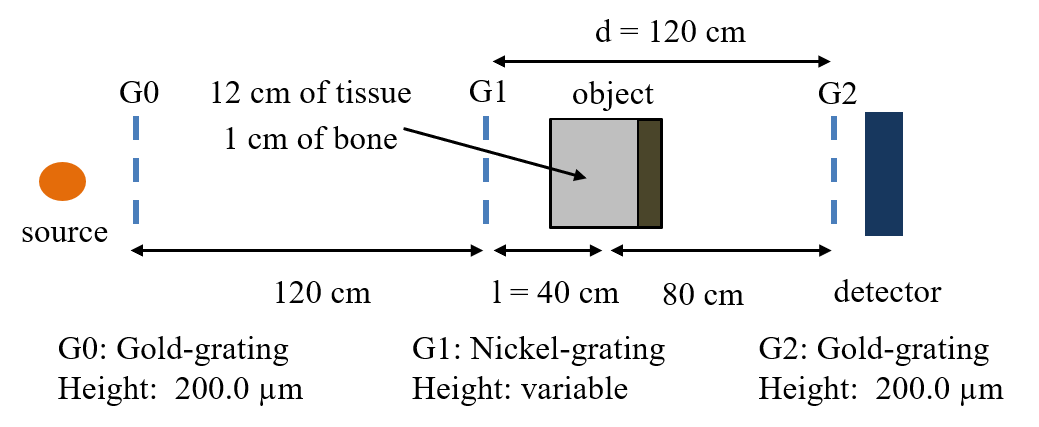} \\  
    \caption{Overview the setup geometry that is kept fixed for all three imaging methods. The three gratings are removed for spectral imaging.} 
    \label{fig:setup_geometry}
\end{figure}

To compare the noise level of the projected electron densities for all three imaging methods and to test the predictions of the noise analysis framework, we simulated a radiography measurement of an homogeneous object. The X-ray beam had to penetrate $12\ \rm{cm}$ of soft tissue and $1\ \rm{cm}$ of cortical bone, \added{which corresponds to a PED of $12 \ \rm{cm} \cdot 3.52 \cdot 10^{23} \ \rm{cm}^{-3} + 1 \  \rm{cm} \cdot 5.95 \cdot 10^{23} \ \rm{cm}^{-3}  = 4.82 \cdot 10^{24} \ \rm{cm}^{-2}$}. \added{The aforementioned thicknesses} could reflect typical path lengths for medical imaging tasks (e.g. thorax radiography or head CT) . The simulated object had no internal microstructure, i.e. it did not generate a dark-field signal. We assumed a tungsten X-ray source and a PCD with a pixel size of $200 \ \rm{\mu m}$, a $2 \ \rm{mm}$ thick cadmium telluride sensor and two thresholds per pixel. As explained in the methods section, only one detector row (containing 400 pixels) was simulated. The detector response was simulated with an empirical model \cite{schlomka2008experimental} that includes sensor effects (charge sharing, K-escapes). The source spectrum and the detector response were discretized in $1 \ \rm{keV}$ steps. For DPC and SDPC imaging, a symmetric three-grating interferometer (operated at the first fractional Talbot order) was inserted into the beam path. The attenuation gratings (G0 and G2) were assumed to be made of gold with a grating height of $200 \ \rm{\mu m}$. The G1 grating was assumed to be made of nickel and to induce a phase shift of $\pi/2$ for the design energy. \added{Although the duty cycle of the gratings influences the noise level \cite{thuering2014performance}, it was kept fixed at its standard value (0.5) in this simulation study in order to reduce the number of possible acquisition parameter combinations.} The total length of the simulated setup was $2.4 \ \rm{m}$ and the object was placed between G1 and G2 ($40 \ \rm{cm}$ away from G1). Figure \ref{fig:setup_geometry} shows an overview of the setup geometry that was kept fixed for all three imaging methods.
All other imaging parameters (acceleration voltage, threshold positions and design energy of the interferometer) were optimized individually for each imaging method (if applicable). For all three imaging methods, the low threshold of the PCD was kept fixed at $15 \ \rm{keV}$. The position of the high threshold was optimized for SDPC and spectral imaging while only one threshold (at $15 \ \rm{keV}$) was used for DPC imaging. For DPC and SDPC imaging, the design energy was tuned by changing the height of the phase shifting grating as well as the grating periods. Changing the grating periods is necessary to keep the first fractional Talbot distance at the detector position. Stepping curves (using 5 steps equally distributed between $0$ and $2\pi$) were simulated with a wave-optical simulation package based on Fresnel propagation \cite{engelhardt2008fractional,wang2010analysis} and the projection approximation \cite{paganin2006coherent}. For all three imaging methods, the variable setup parameters were optimized by minimizing the variance of the PED  predicted by the CRLB while keeping the dose to the object constant ($1 \ \rm{mGy}$). The dose was estimated based on an empirical model that was fitted to Monte-Carlo simulations \cite{boone1992parametrized}. Since the G0 and G1 gratings are placed between the source and the object, they only influence the (spectral) photon flux incident on the object. The G2 grating, however, attenuates photons that have (potentially) contributed to the dose delivered to the object because it is placed between the object and the detector.

\section{Results}
\begin{figure}[h]
    \centering
    %\captionsetup[subfloat]{farskip=0pt,captionskip=0pt}
   \subfloat[]{\includegraphics[trim={0.0cm 0.0cm 0.0cm 0.cm},clip,width = 2.85in]{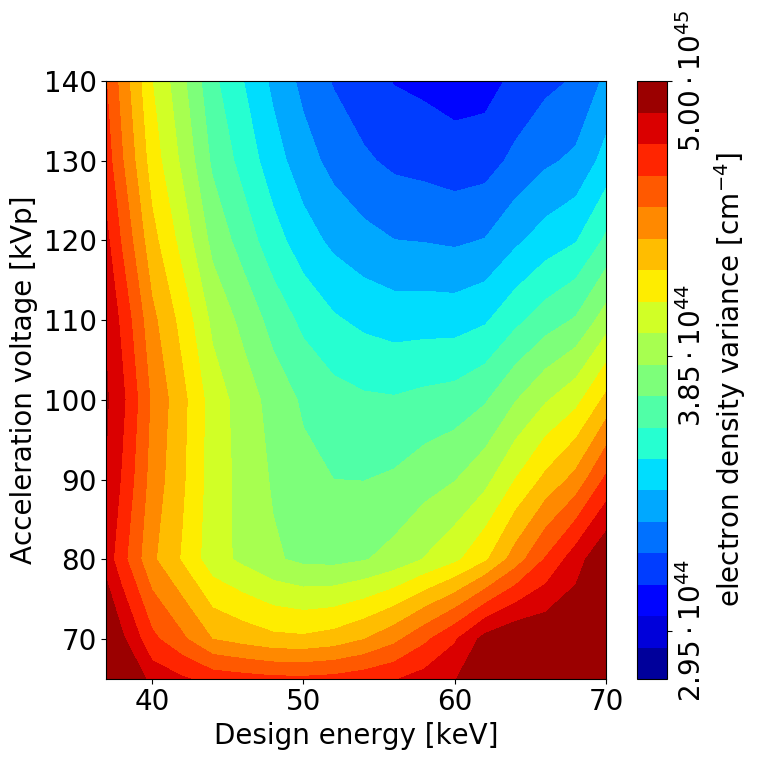}}
   \hspace{0.2in}
   \subfloat[]{\includegraphics[trim={0.0cm 0.0cm 0.0cm 0.cm},clip,width = 2.85in]{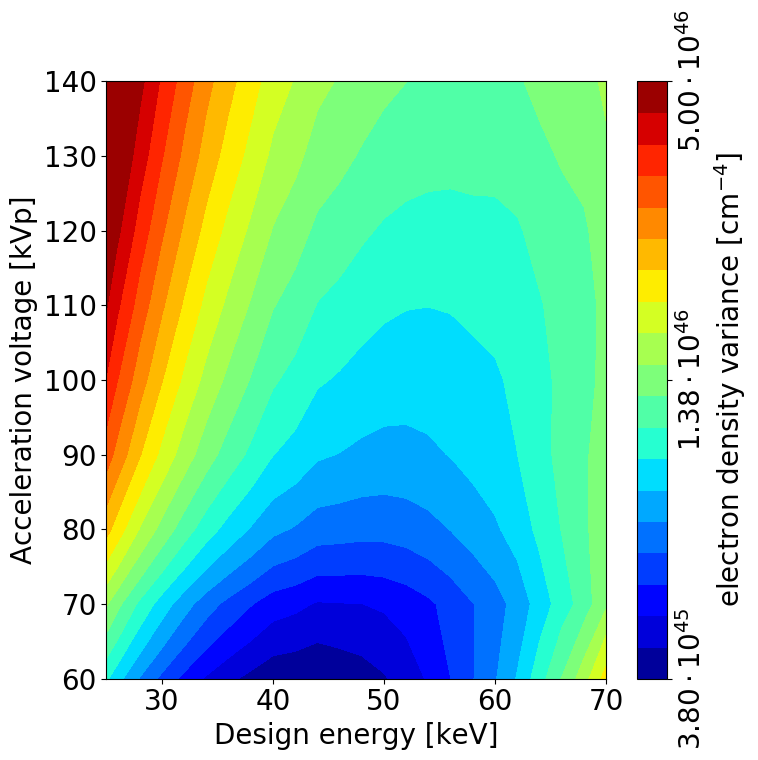}}   
    \caption{(a): Contour plot of the predicted projected electron density (PED) variance (at the optimum threshold position) for SDPC imaging  as a function of the acceleration voltage and the design energy. (b): Predicted PED variance for DPC imaging as a function of the acceleration voltage and the design energy.} 
    \label{fig:crlb_surface}
\end{figure}

\begin{figure}[h]
    \centering
    %\captionsetup[subfloat]{farskip=0pt,captionskip=0pt}
   \subfloat[]{\includegraphics[trim={0.0cm 0.0cm 0.0cm 0.cm},clip,width = 3.0in]{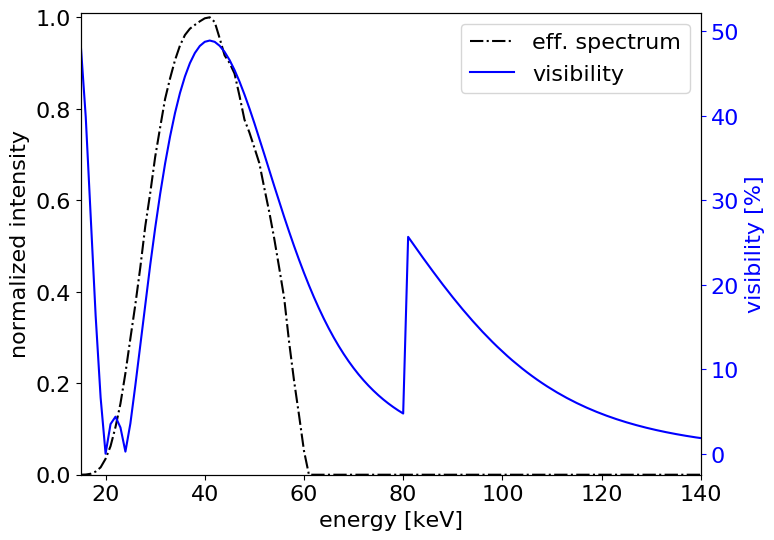}}
   \subfloat[]{\includegraphics[trim={0.0cm 0.0cm 0.0cm 0.cm},clip,width = 3.0in]{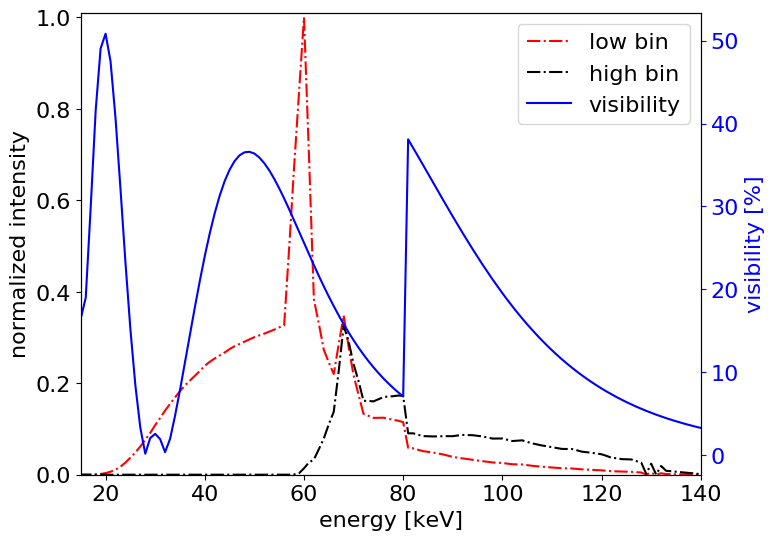}}  
    \caption{Energy-dependent visibilities and effective spectra for DPC (a) and SDPC (b) imaging using the optimum acquisition parameters given in table \ref{tab:opt_params}.} 
    \label{fig:vis_and_spectra}
\end{figure}

\begin{figure}[h]
    \centering
    %\captionsetup[subfloat]{farskip=0pt,captionskip=0pt}
   \subfloat[]{\includegraphics[trim={0.0cm 0.0cm 0.0cm 0.cm},clip,width = 3.0in]{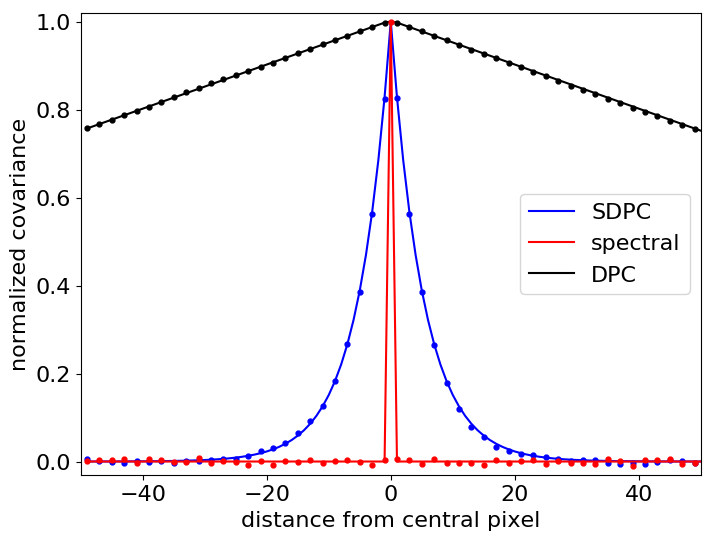}}
   \subfloat[]{\includegraphics[trim={0.0cm 0.0cm 0.0cm 0.cm},clip,width = 3.0in]{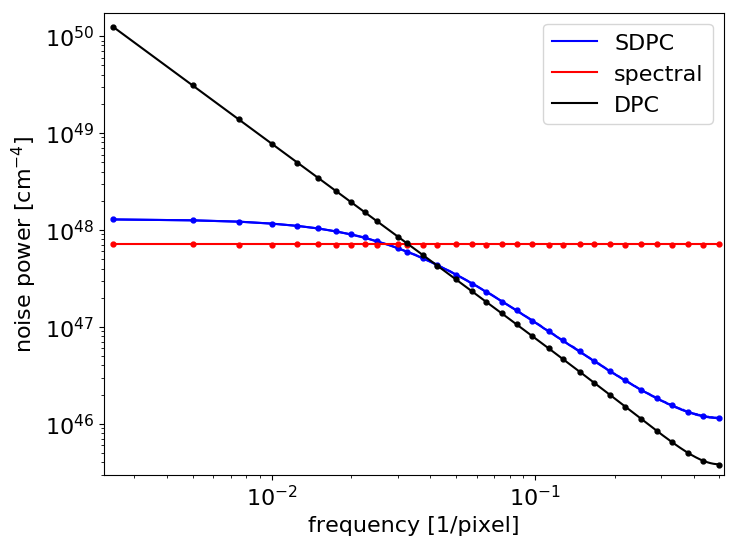}}  
    \caption{(a): Experimentally calculated normalized covariances between the central detector pixel and the other pixels (dots) together with the theoretical predictions (solid lines). (b): Noise power spectra for all three imaging methods obtained in the numerical experiment (dots) and the corresponding theoretical predictions (solid lines). } 
    \label{fig:cov_and_nps}
\end{figure}

\begin{figure}[h]
    \centering
    %\captionsetup[subfloat]{farskip=0pt,captionskip=0pt}
   \subfloat[]{\includegraphics[trim={0.0cm 0.0cm 0.0cm 0.cm},clip,width = 3.0in]{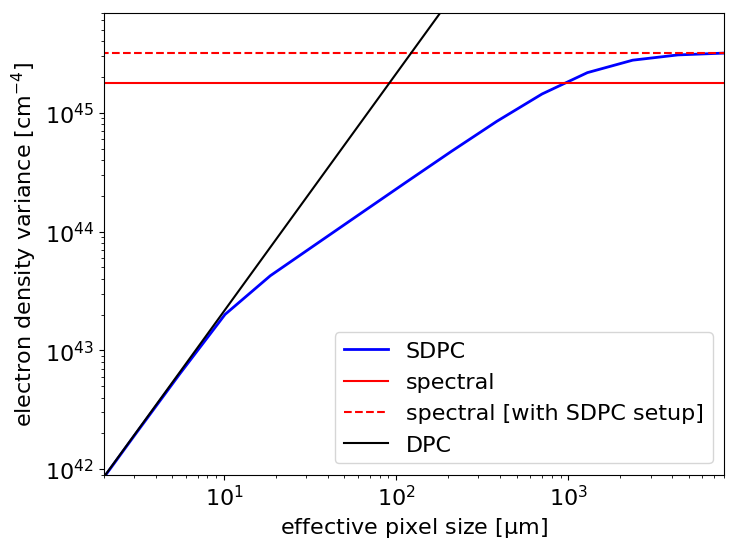}}
   \subfloat[]{\includegraphics[trim={0.0cm 0.0cm 0.0cm 0.cm},clip,width = 3.0in]{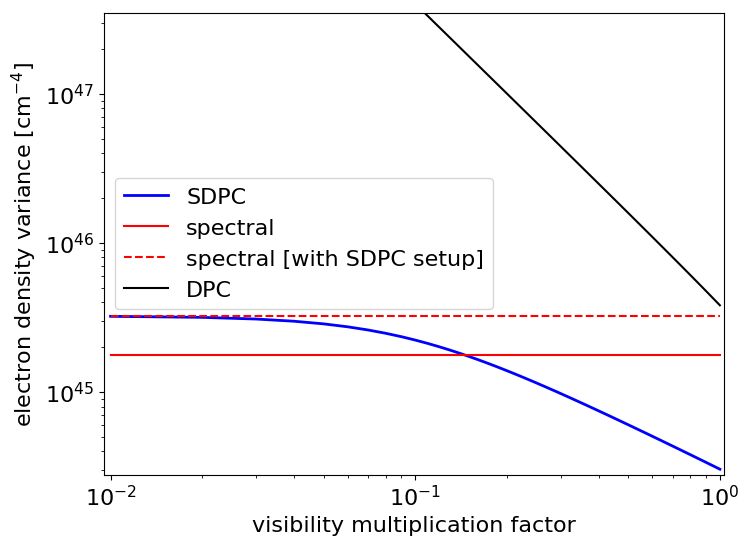}}   
    \caption{(a): Minimum PED variances for all three imaging methods as a function of the effective pixel size. For visualization purposes, the number of photon counts per pixel (instead of the dose delivered to the object) was kept constant.\\
    (b): PED variances with standard resolution and imaging parameters (see table \ref{tab:opt_params}) as a function of the visibility. For this experiment, the energy-dependent visibilities (see figure \ref{fig:vis_and_spectra}) were multiplied by a visibility reduction factor $m$ ($0<m<1$) to simulate a decreased visibility due to experimental effects. }
    \label{fig:crlb_with_res}
\end{figure}

The method for finding the optimum setup parameters is illustrated in figure \ref{fig:crlb_surface} (a) and (b). Figure \ref{fig:crlb_surface} (a) shows a contour plot of the predicted PED variance for SDPC imaging (at the optimum threshold position) as a function of the acceleration voltage and the design energy. \added{As demonstrated in the last section, the PED for the simulated decomposition task is on the order of $10^{24} \ \rm{cm}^{-2}$, while the corresponding standard deviations are approximately two orders of magnitude lower. This explains the larger numerical values for the PED variances of up to $10^{46} \ \rm{cm}^{-4}$.} The minimum variance is achieved for an acceleration voltage of $140 \ \rm{kVp}$ and a design energy of $60 \  \rm{keV}$. As the acceleration voltage is decreased, the optimum design energy also decreases.  Figure \ref{fig:crlb_surface} (b) shows a similar contour plot for DPC imaging. In comparison with SDPC imaging, the optimum acceleration voltage  is much lower (for this particular decomposition task). However, similarly to SDPC imaging, the optimum design energy is positively correlated with the acceleration voltage. This behavior is reasonable because a large part of the spectrum should be concentrated around the design energy of the interferometer for optimum performance. 
Table \ref{tab:opt_params} shows the optimum setup parameters as well as the minimum PED variances for all three imaging methods. Interestingly, the optimum setup parameters for SDPC imaging are very similar to those for spectral imaging, while DPC imaging favors a considerably lower acceleration voltage ($60 \ \rm{kVp}$) and design energy ($44 \ \rm{keV}$). 
\begin{table}
\caption{Optimum imaging parameters and PED variances for spectral, DPC and SDPC imaging. The optimum parameters were determined by minimizing the PED variance predicted by the CRLB while keeping the dose to the object constant ($1 \ \rm{mGy}$).}
\label{tab:opt_params}
\centering
\begin{tabular}{ c| c|c|c|c}
      & optimization range& \multicolumn{3}{c}{optimum parameters} \\
    & & spectral & DPC & SDPC \\
   %variance  & conv. spectral & SDPC  & dose reduction \\ 
  \hline
  acceleration voltage $\rm{[kVp]}$ & $40-140$  & $140 $   & $60 $  &  $140 $ \\
  high threshold $\rm{[keV]}$ & $15-140 $ & $63 $   & - & $65 $ \\
  design energy $\rm{[keV]}$ & $20-70 $ & - & $44$ &$60$  \\
    G2 grating period $[\rm{\mu m}]$  & $9.22 -17.25 $ & - & $11.63$ &$9.96$  \\
  \hline
  min. PED variance $\rm{[cm^{-4}]}$ & &  $1.77\cdot 10^{45}$ & $3.82\cdot 10^{45}$ & $3.04\cdot 10^{44}$
\end{tabular}
\end{table}
A possible explanation for these results can be found by looking at figure \ref{fig:vis_and_spectra}, which shows the energy-dependent visibility as well as the effective spectra for DPC and SDPC imaging with optimum setup parameters. The effective spectrum includes the source spectrum, the attenuation of the gratings and the detector response. Kottler et. al \cite{kottler2010dual} demonstrated that a grating interferometer with a $\pi/2$ phase shifting grating (G1) exhibits a second visibility peak at twice the design energy. The second important factor that influences the visibility is the attenuation of the gold gratings. In the range from $60-80 \, \rm{keV}$, the gold gratings become increasingly transparent, but for energies above the K-edge of gold ($80.7 \, \rm{keV}$), the attenuation of the gratings and thus also the visibility rises sharply.
In the case of DPC imaging, a lower effective energy (see equation \ref{E_eff_dpc}) is preferable  because it increases the sensitivity of the setup (compare equation \ref{delta_phi_from_rhoe_dpc}). In other words, a fixed gradient of the PED causes a larger phase shift for lower effective energies. Since the spatial resolution and the setup geometry are fixed for our simulation, the sensitivity can only be tuned via the design energy and the effective energy. The acceleration voltage ($60 \ \rm{kVp}$) is matched to the design energy ($44 \ \rm{keV}$) so that a large fraction of the effective spectrum is concentrated around the visibility peak. Achieving a high  visibility is important  because for DPC imaging, the variance of the PED is proportional to the visibility squared (compare equation \ref{sigma_delta_phi_interp}).
The comparatively high acceleration voltage ($140 \ \rm{kVp}$) for SDPC imaging leads to a decreased sensitivity compared with DPC imaging. However, this decrease in sensitivity is compensated by the improved performance of spectral imaging at higher acceleration voltages. Moreover, the decrease in sensitivity and visibility compared with DPC imaging is less pronounced for the low energy bin. Due to the higher acceleration voltage, the optimum design energy for SDPC imaging is considerably larger ($60 \ \rm{keV}$). Consequently, the visibility peak for low energies is reduced compared with DPC imaging. On the other hand, the visibility for energies larger than the K-edge of Gold ($80.7 \, \rm{keV}$) is increased.\\
The predicted PED variances show that for this particular imaging task, SDPC imaging can achieve considerably lower noise levels compared with spectral and DPC imaging (variance reduction by a factor of 5.8 and 12.6. respectively). It is important to note that  for DPC imaging, the variance depends approximately linearly on the number of pixels in one detector row. This behavior is caused by the integration that is required to convert the differential phase shifts to a PED profile (see equation \ref{dpc_int_lr}). For SDPC imaging, the PED variance only depends very weakly on the number of pixels. The reason for this weak dependence will be explained later when we analyze the noise correlations for all three imaging methods. \\
\begin{figure}[h]
    \centering
    %\captionsetup[subfloat]{farskip=0pt,captionskip=0pt}
   \includegraphics[trim={0.0cm 0.0cm 0.0cm 0.cm},clip,width = 3.6in]{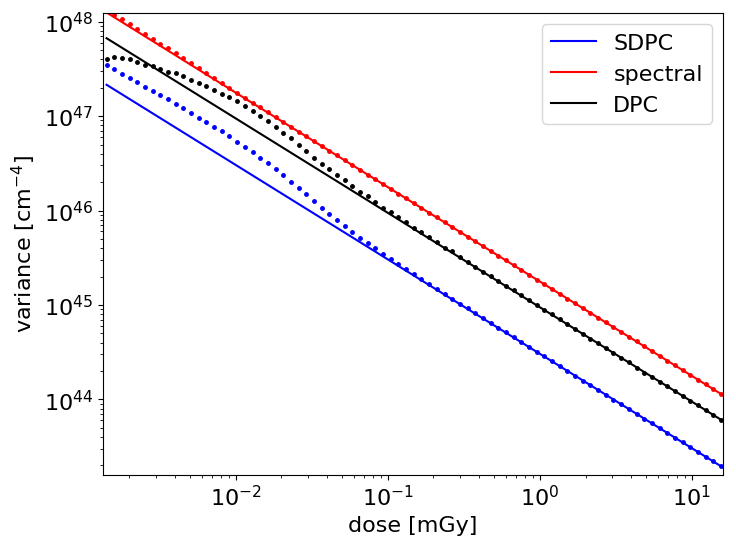} \\  
    \caption{Experimentally obtained PED variances (dots) and theoretical predictions (solid lines) for all three imaging methods as a function of the dose level.} 
    \label{fig:ml_noise_sweep}
\end{figure}
We conducted numerical experiments with 100 different dose levels ranging from $0.001$ to $10 \ \rm{mGy}$ in order to test if the ML-based estimators achieves the predicted PED variances. For these experiments, we reduced the detector size to 100 pixels to reduce the computational time. For each dose level and imaging method, $10,000$ different noise realizations were simulated and the ML-based estimators (compare equations \ref{ml_spectral},\ref{ml_dpc},\ref{ml_sdpc}), were used to calculate the PED. Figure \ref{fig:ml_noise_sweep} shows the experimentally achieved PED variances (dots) together with the variances predicted by the CRLB (solid lines) for all three imaging methods. All estimators achieve the CRLB for a large range of dose levels ($\approx 0.1 -10 \  \rm{mGy}$). For lower dose levels, deviations from the CRLB are observable, especially for DPC and SDPC imaging. In the case of DPC imaging, a similar dependency of the variance on the dose level (or photon statistics) has been reported for a Fourier-based estimator \cite{raupach2011analytical}. In this simulation, DPC imaging achieves a lower variance than spectral imaging because of the reduced detector size. However, the main goal of this simulation was to test the predictions of the noise analysis framework rather than comparing the performance of the three imaging methods. \\
It has already been pointed out that the integration step for DPC imaging causes long-range correlations between the PED values for one detector row. Consequently, the NPS for DPC imaging is dominated by low frequencies. In  the case of spectral imaging, the basis material line integrals can be determined separately for each detector pixel and thus the PED values determined by spectral imaging are uncorrelated (at least for an ideal PCD). Therefore, the noise power should be equally distributed between all frequencies (``white noise"). Since SDPC imaging is a combination of both aforementioned imaging methods, it could be expected that the NPS for SDPC imaging is a mixture between the noise power spectra for the other two imaging methods.
In section \ref{sec:cov_and_nps_from_crlb}, we have presented a framework for predicting the PED covariances as well as the noise power spectra for all three imaging methods. We conducted another numerical experiment to verify the predicted covariances and noise power spectra. In this simulation, the dose was fixed at $1 \ \rm{mGy}$ and the original detector size (400 pixels) was used. For each of the three imaging methods, 50,000 different noise realizations were simulated and the PEDs were calculated separately for each noise realization. Figure \ref{fig:cov_and_nps} (a) shows the experimentally calculated normalized covariances (dots) together with the theoretical predictions (solid lines) for all three imaging methods. More precisely, the covariances between the PED for the central detector pixel (i.e. pixel index 200) and the PEDs of all other pixels is shown. In figure \ref{fig:cov_and_nps} (b), the experimentally obtained noise power spectra are plotted together with the theoretical predictions. We used equation \ref{app:def_nps} for the experimental calculation of the NPS.
For both the covariances and the noise power spectra, the theoretical predictions are in good agreement with the experimentally obtained values. As expected, the covariance of the PED between two different detector pixels is zero for spectral imaging. For DPC imaging, the covariance decreases linearly with the distance from the central detector pixel until it drops to zero at the edges of the detector (not shown in figure \ref{fig:cov_and_nps} (a)). This reflects the well-known long-range noise correlations introduced by the integration step. In the case of SDPC imaging, however, the covariance rapidly drops to zero with increasing distance from the central pixel. For the imaging parameters given in table \ref{tab:opt_params}, almost no noise correlations are observable if the distance to the central pixel is larger than  approximately $25$ pixels. Compared to DPC imaging, the additional spectral information eliminates long-range noise correlations. On the other hand, it could be argued that compared with spectral imaging, the additional phase shift term for SDPC imaging couples the PED values of neighboring pixels and thus introduces local noise correlations. The covariance graph for SDPC imaging explains the aforementioned weak dependence of the PED variance on the number of detector pixels. Contrary to DPC imaging, the covariance rapidly decreases with the distance between two pixels and thus the assumption of fixed, known PED values at the edges of the detector does not influence the noise level (except for pixels close to the edges). The different noise correlations for the three imaging methods are also reflected in the different noise power spectra. As expected, the noise power is independent of the frequency for spectral imaging, whereas the noise power for DPC imaging increases drastically for lower spatial frequencies. For high frequencies, the noise power spectrum for SDPC imaging is similar to DPC imaging. For low frequencies, however, the noise power does not increase but converges to a constant value that lies slightly above the noise power graph for spectral imaging. The higher noise power for SDPC imaging in the low frequency area can be explained by the attenuation of the G2-grating, which removes a part of the X-ray beam that has contributed to the dose delivered to the object. \\
In our simulation study, we assumed a fixed total length of the setup and a symmetric grating interferometer that operates in the first fractional Talbot distance.
Given these conditions, the effective pixel size is the most important tuning factor for the sensitivity of a DPC or SDPC imaging setup (compare equation \ref{sens_dpc_def} and \ref{delta_phi_from_rhoe_sdpc}). Since the PED variance strongly depends on the sensitivity, we investigated the theoretically predicted performance of all three imaging methods as a function of the effective pixel size. For each pixel size, the optimum imaging parameters were determined with the noise analysis framework presented in the methods section. Figure \ref{fig:crlb_with_res} (a) shows the predicted PED variances as a function of the effective pixel size. For visualization purposes, the number of photon counts per pixel (instead of the dose delivered to the object) was kept constant. This means that the dose delivered to the object was chosen inversely proportional to the squared effective pixel size (i.e. the effective pixel area). If the dose was kept constant, the PED variances  would have to be multiplied by an additional factor that is inversely proportional to the effective pixel area for all three imaging methods.
Under these conditions, the PED variance for spectral imaging is independent of the pixel size. In the case of DPC imaging, the variance is proportional to the squared effective pixel size, which reflects the well-known sensitivity-dependent noise level for DPC imaging (compare also equation \ref{error_prop_delta_phi_rhoe}). For large effective pixel sizes ($a > 4 \ \rm{mm}$), the PED variance for SDPC imaging is almost constant and approximately twice as large as the variance for spectral imaging. It coincides with the PED variance one would obtain if spectral imaging was performed with the ``unnecessary" grating interferometer inserted into the beam path (see dashed line in figure \ref{fig:crlb_with_res} (a)). It follows that for larger pixel sizes (or low sensitivities), the additional information on the basis material line integrals provided by the phase shift of the stepping curve can be neglected compared to the spectral information.  Consequently, SDPC imaging reduces to spectral imaging in the limit of large effective pixel sizes. For this particular imaging task, SDPC imaging and spectral imaging have equal PED variances for an effective pixel size of $a = 950  \ \rm{\mu m}$. At this point, the benefit of the additional phase shift information is counterbalanced by the reduced photon statistics caused by the attenuation of the G2 grating. For very small pixel sizes ($a < 10 \ \rm{\mu m}$), the SDPC variances almost coincide with the DPC variances. In this case, SDPC imaging reduces to DPC imaging because the spectral information provides little additional value for determining the  PED in comparison to the phase shift of the stepping curve. In the area between these limiting cases ($10 \ \rm{\mu m} < a < 950 \ \rm{\mu m}$), the SDPC variance lies below both the spectral and the DPC variance because both the spectral and the phase shift information are relevant for determining the PED. The largest variance reduction (by a factor of 8.8) compared to both DPC and spectral imaging is achieved for the point of equal variance of spectral and DPC imaging ($a= 90 \ \rm{\mu m}$). At this point, the spectral and the phase shift information are of equal importance for determining the PED.\\ It is important to note that the predicted PED variance only depends on the geometrical imaging parameters (pixel size, object position, grating distances, grating periods) via the sensitivity. Consequently, similar trends for the PED variances could have been observed by varying geometrical parameters other than the pixel size. \added{However, increasing the sensitivity via these geometrical parameters also increases the phase shift of the stepping curve that a given sample causes (via the gradient of the PED). For DPC imaging, this might lead to additional phase wrapping artifacts. As will be discussed in the next section, phase wrapping artifacts can in theory be avoided for SDPC imaging.}  \\
The number of detector pixels ($N= 400$) was kept constant in this analysis of the noise characteristics. This means that the field of view decreases with decreasing effective pixel size $a$. If the field of view was kept constant, the total number of detector pixels would have to be scaled by $1/a$. As discussed in the results section, increasing the number of detector pixels would have an adverse effect on the PED variance for DPC imaging, but the influence on SDPC imaging would be negligible as long as the spectral information prevents long-range noise correlations.\\
In a real experiment, various undesirable effects (e.g. vibrations, grating imperfections) could cause a reduced visibility of the stepping curves compared with the simulations. Moreover, we have assumed that the object does not generate a dark-field signal which would also reduce the effective visibility. We therefore investigated the influence of a visibility reduction on the PED variance for SDPC and DPC imaging. For this study, the imaging parameters  were kept fixed at their optimum values (compare table \ref{tab:opt_params}). We simulated a visibility reduction by multiplying the energy-dependent visibilities (compare figure \ref{fig:vis_and_spectra}) with a visibility reduction parameter $m$ ($0<m<1$).  Figure \ref{fig:crlb_with_res} (b) shows the predicted PED variances for SDPC and DPC imaging as a function of the visibility reduction parameter. For comparison, the variances for spectral imaging with and without the interferometer in the beam path are also plotted as constant lines. As expected, the variance for DPC imaging rises rapidly with decreasing visibility ($ \sigma^2(\rho_{\rm{e}}) \propto \left(Ve^{-\epsilon}\right)^{-2}$, compare equation \ref{sigma_delta_phi_interp}). In the case of SDPC imaging, however, the variances rises much more slowly with decreasing visibility (approximately proportionally to $\left(Ve^{-\epsilon}\right)^{-1}$)  before it approaches a constant value for low visibilities. This is reasonable because in the limit of low visibilities, SDPC imaging effectively reduces to spectral imaging (with the gratings in the beam path). For $m=0.15$, the variances for SDPC imaging and spectral imaging are equal. This means that for this particular reconstruction task, the visibility can be reduced by a factor of $1/m = 6.7$ compared with the idealized simulation before the attenuation of the G2 grating outweighs the variance reduction achieved by the additional phase shift information.

\section{Discussion}
A quantitative comparison between conventional attenuation imaging and DPC imaging is difficult because of the different underlying contrast generating mechanisms for these two imaging methods. For example, a contrast-to-noise ratio comparison \cite{herzen2009quantitative,zambelli2010radiation} is complicated because the differences in contrast for the two imaging methods depends on the choice of the investigated materials. Moreover, depending on the setup parameters, the contrast between different materials can vanish completely for both imaging methods. For this reason, we concentrated on electron density images, which can be calculated with all three imaging methods (spectral, DPC and SDPC imaging) under consideration in this comparison study. Nevertheless, the presented noise analysis framework can also be used to determine the optimum acquisition parameters for other imaging tasks, such as dark-field imaging or basis material decomposition. Although we have focused on projection imaging, the noise analysis framework could be generalized to computed tomography imaging (e.g. by using error propagation for the filtered backprojection \cite{avinash1988principles}). \\
For the investigated imaging task, our analysis predicts highly reduced noise levels for SDPC imaging in comparison to both DPC imaging and spectral imaging. To achieve a fair comparison, the imaging parameters were optimized individually for each of the three methods. For a large range of clinically realistic dose levels, the ML-based estimators achieve the predicted noise levels for all three imaging methods. The validity of the predicted covariances and noise power spectra was confirmed by additional numerical simulations. Our analysis focused on the projected electron density, but earlier studies have shown that the noise advantage of SDPC imaging compared to spectral imaging  is also transferred to the basis material images \cite{mechlem2019spectral}. In our study, DPC imaging only outperforms spectral imaging for comparatively high spatial resolutions ($a< 90 \ \rm{\mu m}$), however this conclusion does not apply to SDPC imaging. For a large range of clinically relevant effective pixel sizes (up to $950 \ \rm{\mu m}$), SDPC imaging theoretically outperforms the other two imaging methods by simultaneously using both the spectral and phase contrast information. Although we have only considered one imaging scenario in this work, the general trends should apply to a large range of imaging tasks. Nevertheless, the location of the break-even points between the three imaging methods will vary depending on the object size and the chemical composition. In a real experiment, both grating imperfections and ultra-small angle scattering by the object (i.e. nonzero dark-field signals) could reduce the visibility of the stepping curves compared to our simulations. The noise reduction for SDPC imaging compared to spectral imaging will thus be smaller than theoretically predicted. Nonetheless, because of the large theoretical noise reduction factors, we believe that SDPC imaging can also outperform the other two imaging methods in real experiments. Moreover, the noise level for SDPC imaging increases less rapidly with decreasing visibility compared with DPC imaging (see figure \ref{fig:crlb_with_res} (b)).  \\
Regardless of the potential for noise reduction, the combination of spectral and phase-contrast imaging provides additional information that is inaccessible with the individual imaging methods. Compared with spectral imaging, the dark-field image yields additional information about the object's microstructure and, compared with DPC imaging, two basis material images can be calculated and beam hardening artifacts in all three imaging channels can be corrected. \\
Raupach and Flohr \cite{raupach2012performance} state that the noise correlations for DPC imaging are disadvantageous for clinical diagnosis. They argue that the dominance of low frequencies in the NPS is unfavourable for recognizing structures. Furthermore, because of the low-frequency noise, reducing the noise level by pixel binning is less effective \added{compared with conventional attenuation-based imaging}. In general, these arguments do not apply to SDPC imaging because of the fundamentally different noise correlations compared to DPC imaging. An exception is the limit of very small pixel sizes (or high sensitivities), where SDPC imaging effectively converges to DPC imaging. In our study, the influence of the spectral information causes a rapid decrease of the covariance between two pixels with increasing distance. Consequently, there is no excessive low frequency noise.
Furthermore, Raupach and Flohr \cite{raupach2011analytical} argue that - contrary to conventional attenuation imaging - there is a dose limit for DPC imaging below which no meaningful signals can be extracted. The interferometric measurement acquisition process only allows unique determination of the phase shift of the stepping curve across a $2 \pi$ interval. Larger phase shifts are wrapped back into this interval. According to Raupach and Flohr, statistical phase wrapping leads to a loss of the phase shift information for low dose levels. More precisely, the average extracted phase shift is biased to zero, regardless of the underlying physical phase shift if a standard averaging procedure is used. \deleted{However, this bias can be avoided by using a circular average that considers the underlying $2 \pi$ periodicity of the phase shifts. Nevertheless, standard averaging methods can be used for SDPC imaging.} As a result of the additional spectral information when compared to DPC imaging, the calculated phase shift is not restricted to a $2 \pi$ interval \added{for SDPC imaging}. Qualitatively speaking, the spectral information determines in which $2 \pi$ interval the phase shift lies (via the PED profile) and the exact value is fine-tuned by the stepping curve information. In principle, there is thus no information loss through statistical phase wrapping. However, it has been shown that the log-likelihood function for SDPC imaging has local optima that can be explained by an analogy to the phase wrapping problem for DPC imaging \cite{mechlem2019spectral}. Although we could avoid the convergence to local optima in previous simulation studies by choosing a suitable initial guess, this strategy is likely to fail at extremely low dose levels. Nevertheless, the convergence to local optima is an optimization problem rather than a fundamental restriction of SDPC imaging because it could in principle be avoided with a global optimization strategy (or by incorporating prior information in the form of a regularization term). \\

\section{Conclusion}
In this work, we have developed a noise analysis framework that allows the calculation of (co-) variances and noise power spectra for spectral, DPC and SDPC imaging. An important practical application of this framework is finding the optimum imaging parameters for all three methods. Our analysis shows that the combination of spectral and phase-contrast imaging is a promising imaging technique with various advantages compared with the individual methods. SDPC imaging provides additional information compared with both spectral imaging (dark-field image) and DPC imaging (basis material images). Moreover, we demonstrated that SDPC imaging enables a strong noise (or dose) reduction compared with the other two imaging methods for a large range of clinically relevant pixel sizes. Finally, the additional spectral information compared to DPC imaging eliminates excessive low-frequency noise, which can be a major drawback of DPC imaging, especially in projection space.

\section{Appendix}
\subsection{Noise propagation for the integration step in DPC imaging}
Applying standard error propagation techniques to equation \ref{dpc_int_l_to_r}, the variance of the PED for a certain pixel ($\sigma^2 (\tilde{\rho}_e^i)$) can be calculated:
\begin{equation}
\sigma^2 (\rho_{\rm{e}}^i) = \frac{1}{\mathcal{S}^2} \sum_{q=1}^{i-1} \sigma^2(\Delta \phi_q),
\end{equation}
where $\sigma^2(\Delta \phi_q)$ is the variance of the differential phase shift (which can be calculated using the CRLB). In the special case of a homogeneous sample (i.e. $\sigma^2(\Delta \phi_q) = \text{const.} =  \sigma^2(\Delta \phi) \  \forall q$), the average variance of the electron density $\sigma_{\rm{avg}}^2(\rho_{\rm{e}})$ can be calculated as:
\begin{equation}
\label{var_int_l}
\begin{split}
&\sigma_{\rm{avg}}^2(\rho_{\rm{e}}) = \frac{1}{N-2} \sum_{i=2}^{N-1} \sigma^2 (\rho_{\rm{e}}^i) = \frac{1}{\mathcal{S}^2(N-2)} \sum_{i=2}^{N-1} \sum_{q=1}^{i-1} \sigma^2(\Delta \phi)  \\
&= \frac{1}{\mathcal{S}^2(N-2)} \sum_{i=2}^{N-1} (i-1) \sigma^2(\Delta \phi)  =  \frac{N-1}{2\mathcal{S}^2}\sigma^2(\Delta \phi).
\end{split}
\end{equation}
The pixels with index $1$ and $N$ are excluded from the average variance calculation because it was assumed that the PED is known with certainty at the edges of the detector. The same result is obtained if the integration from right to left (equation \ref{dpc_int_r_to_l}) is considered.
If the left- and right-integrated  PED profiles are averaged (see equation \ref{dpc_int_lr}), the variance of the PED is given by:
\begin{equation}
\label{var_int_lr}
\sigma^2 (\rho_{\rm{e}}^i) = \frac{1}{4\mathcal{S}^2}\left( \sum_{q=1}^{i-1} \sigma^2(\Delta \phi_q) + \sum_{q=i}^{N-1} \sigma^2(\Delta \phi_q) \right) = \frac{1}{4\mathcal{S}^2} \sum_{q=1}^{N-1} \sigma^2(\Delta \phi_q).
\end{equation}
As can be seen from equation \ref{var_int_lr}, the variance of the electron density is independent of the detector pixel index. Consequently, the variance is spatially constant, even for a nonhomogeneous sample. In the special case of a homogeneous sample, $\sigma_{\rm{avg}}^2(\rho_{\rm{e}})$ is given by:
\begin{equation}
\sigma_{\rm{avg}}^2(\rho_{\rm{e}}) = \frac{1}{N-2} \sum_{i=2}^{N-1} \sigma^2 (\rho_{\rm{e}}^i) = \frac{1}{4\mathcal{S}^2} (N-1) \sigma^2(\Delta \phi).
\end{equation}
Comparing equation (\ref{var_int_lr}) and (\ref{var_int_l}), it follows that (in the case of a homogeneous sample) the average variance is reduced by a factor of two when averaging left- and right-integrated electron density profiles. 
\subsection{Calculating the noise power spectrum from the covariances}
In this section, we derive how the noise power spectrum (NPS) can be calculated from an estimate of the covariances (compare equation \ref{nps_from_cov}).
For a homogeneous object, the NPS can be calculated as \cite{siewerdsen2002framework}:
\begin{equation}
\label{app:def_nps}
{\rm{NPS}}(k) = E \left[ {\left| \mathcal{F} ({\bar{\rho}_{\rm{e}}(x)}  \right|}^2 \right], \quad \bar{\rho}_{\rm{e}}(x) = \rho_{\rm{e}}(x) - E \left[\rho_{\rm{e}}(x) \right],
\end{equation}
where $k$ is the spatial frequency and $\bar{\rho}_e(x)$ is the offset-corrected PED as a function of the spatial coordinate $x$. In discrete form, the NPS is given by:
\begin{equation}
\begin{split}
{\rm{NPS}}(k) &= E \left[ {\left| \sum_{q=0}^{N-1} e^{-2\pi j \frac{qk}{N}} \bar{\rho}_{\rm{e}}^{(q+1)} \right|}^2 \right] = E \left[  \left(\sum_{q=0}^{N-1} e^{-2\pi j \frac{qk}{N}} \bar{\rho}_{\rm{e}}^{(q+1)}\right) \left( \sum_{u=0}^{N-1} e^{+2\pi j \frac{uk}{N}} \bar{\rho}_{\rm{e}}^{(u+1)} \right)  \right] \\
&= E \left[  \sum_{q=0}^{N-1} \sum_{u=0}^{N-1} e^{2\pi j \frac{(u-q)k}{N}} \bar{\rho}_{\rm{e}}^{(q+1)} \bar{\rho}_{\rm{e}}^{(u+1)}  \right] =   \sum_{q=0}^{N-1} \sum_{u=0}^{N-1} e^{2\pi j \frac{(u-q)k}{N}} E \left[\bar{\rho}_{\rm{e}}^{(q+1)} \bar{\rho}_{\rm{e}}^{(u+1)}  \right].
\end{split}
\end{equation}
Since E $\left[\bar{\rho}_{\rm{e}}^{(q+1)} \bar{\rho}_{\rm{e}}^{(u+1)}  \right]$ = ${\rm{Cov}} \left(\rho_{\rm{e}}^{(q+1)}, \rho_{\rm{e}}^{(u+1)} \right)$, the result of equation \ref{nps_from_cov} is obtained. \\

\bibliography{references}

\begin{thebibliography}{10}

\bibitem{Ballabriga2016}
R~Ballabriga, J~Alozy, M~Campbell, E~Frojdh, EHM Heijne, T~Koenig, X~Llopart,
  J~Marchal, D~Pennicard, T~Poikela, et~al.
\newblock Review of hybrid pixel detector readout asics for spectroscopic x-ray
  imaging.
\newblock {\em Journal of Instrumentation}, 11(01):P01007, 2016.

\bibitem{willemink2018photon}
Martin~J Willemink, Mats Persson, Amir Pourmorteza, Norbert~J Pelc, and Dominik
  Fleischmann.
\newblock Photon-counting ct: technical principles and clinical prospects.
\newblock {\em Radiology}, 289(2):293--312, 2018.

\bibitem{mccollough2015dual}
Cynthia~H McCollough, Shuai Leng, Lifeng Yu, and Joel~G Fletcher.
\newblock Dual-and multi-energy ct: principles, technical approaches, and
  clinical applications.
\newblock {\em Radiology}, 276(3):637--653, 2015.

\bibitem{nadjiri2018spectral}
Jonathan Nadjiri, Daniela Pfeiffer, Alexandra~S Straeter, Peter~B No{\"e}l,
  Alexander Fingerle, Hans-Henning Eckstein, Karl-Ludwig Laugwitz, Ernst~J
  Rummeny, Rickmer Braren, and Michael Rasper.
\newblock Spectral computed tomography angiography with a gadolinium-based
  contrast agent.
\newblock {\em Journal of thoracic imaging}, 33(4):246--253, 2018.

\bibitem{symons2018photon}
Rolf Symons, Daniel~S Reich, Mohammadhadi Bagheri, Tyler~E Cork, Bernhard
  Krauss, Stefan Ulzheimer, Steffen Kappler, David~A Bluemke, and Amir
  Pourmorteza.
\newblock Photon-counting computed tomography for vascular imaging of the head
  and neck: first in vivo human results.
\newblock {\em Investigative radiology}, 53(3):135--142, 2018.

\bibitem{schwaiger2018three}
Benedikt~J Schwaiger, Alexandra~S Gersing, Johannes Hammel, Kai Mei, Felix~K
  Kopp, Jan~S Kirschke, Ernst~J Rummeny, Klaus W{\"o}rtler, Thomas Baum, and
  Peter~B No{\"e}l.
\newblock Three-material decomposition with dual-layer spectral ct compared to
  mri for the detection of bone marrow edema in patients with acute vertebral
  fractures.
\newblock {\em Skeletal radiology}, pages 1--8, 2018.

\bibitem{mechlem2018spectral}
Korbinian Mechlem, Thorsten Sellerer, Sebastian Ehn, Daniela M{\"u}nzel, Eva
  Braig, Julia Herzen, Peter~B No{\"e}l, and Franz Pfeiffer.
\newblock Spectral angiography material decomposition using an empirical
  forward model and a dictionary-based regularization.
\newblock {\em IEEE Transactions on Medical Imaging}, 2018.

\bibitem{symons2017photon}
Rolf Symons, Bernhard Krauss, Pooyan Sahbaee, Tyler~E Cork, Manu~N Lakshmanan,
  David~A Bluemke, and Amir Pourmorteza.
\newblock Photon-counting ct for simultaneous imaging of multiple contrast
  agents in the abdomen: an in vivo study.
\newblock {\em Medical physics}, 44(10):5120--5127, 2017.

\bibitem{dangelmaier2018experimental}
Julia Dangelmaier, Daniel Bar-Ness, Heiner Daerr, Daniela Muenzel, Salim
  Si-Mohamed, Sebastian Ehn, Alexander~A Fingerle, Melanie~A Kimm, Felix~K
  Kopp, Loic Boussel, et~al.
\newblock Experimental feasibility of spectral photon-counting computed
  tomography with two contrast agents for the detection of endoleaks following
  endovascular aortic repair.
\newblock {\em European radiology}, pages 1--8, 2018.

\bibitem{pfeiffer2006phase}
Franz Pfeiffer, Timm Weitkamp, Oliver Bunk, and Christian David.
\newblock Phase retrieval and differential phase-contrast imaging with
  low-brilliance x-ray sources.
\newblock {\em Nature physics}, 2(4):258, 2006.

\bibitem{pfeiffer2007hard}
Franz Pfeiffer, Christian Kottler, O~Bunk, and C~David.
\newblock Hard x-ray phase tomography with low-brilliance sources.
\newblock {\em Physical review letters}, 98(10):108105, 2007.

\bibitem{pfeiffer2008hard}
Franz Pfeiffer, Martin Bech, Oliver Bunk, Philipp Kraft, Eric~F Eikenberry,
  Ch~Br{\"o}nnimann, Christian Gr{\"u}nzweig, and Christian David.
\newblock Hard-x-ray dark-field imaging using a grating interferometer.
\newblock {\em Nature materials}, 7(2):134, 2008.

\bibitem{yashiro2010origin}
W~Yashiro, Y~Terui, K~Kawabata, and A~Momose.
\newblock On the origin of visibility contrast in x-ray talbot interferometry.
\newblock {\em Optics express}, 18(16):16890--16901, 2010.

\bibitem{strobl2014general}
Markus Strobl.
\newblock General solution for quantitative dark-field contrast imaging with
  grating interferometers.
\newblock {\em Scientific reports}, 4:7243, 2014.

\bibitem{yashiro2019probing}
Wataru Yashiro, Susumu Ikeda, Yasuo Wada, Kentaro Totsu, Yoshio Suzuki, and
  Akihisa Takeuchi.
\newblock probing surface morphology using x-ray grating interferometry.
\newblock {\em Scientific reports}, 9(1):1--8, 2019.

\bibitem{herzen2009quantitative}
Julia Herzen, Tilman Donath, Franz Pfeiffer, Oliver Bunk, Celestino Padeste,
  Felix Beckmann, Andreas Schreyer, and Christian David.
\newblock Quantitative phase-contrast tomography of a liquid phantom using a
  conventional x-ray tube source.
\newblock {\em Optics express}, 17(12):10010--10018, 2009.

\bibitem{zambelli2010radiation}
Joseph Zambelli, Nicholas Bevins, Zhihua Qi, and Guang-Hong Chen.
\newblock Radiation dose efficiency comparison between differential phase
  contrast ct and conventional absorption ct.
\newblock {\em Medical physics}, 37(6Part1):2473--2479, 2010.

\bibitem{donath2010toward}
Tilman Donath, Franz Pfeiffer, Oliver Bunk, Christian Gr{\"u}nzweig, Eckhard
  Hempel, Stefan Popescu, Peter Vock, and Christian David.
\newblock Toward clinical x-ray phase-contrast ct: demonstration of enhanced
  soft-tissue contrast in human specimen.
\newblock {\em Investigative radiology}, 45(7):445--452, 2010.

\bibitem{raupach2011analytical}
Rainer Raupach and Thomas~G Flohr.
\newblock Analytical evaluation of the signal and noise propagation in x-ray
  differential phase-contrast computed tomography.
\newblock {\em Physics in Medicine \& Biology}, 56(7):2219, 2011.

\bibitem{raupach2012performance}
Rainer Raupach and Thomas Flohr.
\newblock Performance evaluation of x-ray differential phase contrast computed
  tomography (pct) with respect to medical imaging.
\newblock {\em Medical physics}, 39(8):4761--4774, 2012.

\bibitem{braig2018direct}
Eva Braig, Jessica B{\"o}hm, Martin Dierolf, Christoph Jud, Benedikt
  G{\"u}nther, Korbinian Mechlem, Sebastian Allner, Thorsten Sellerer, Klaus
  Achterhold, Bernhard Gleich, et~al.
\newblock Direct quantitative material decomposition employing grating-based
  x-ray phase-contrast ct.
\newblock {\em Scientific reports}, 8(1):16394, 2018.

\bibitem{mechlem2019spectral}
Korbinian Mechlem, Thorsten Sellerer, Manuel Viermetz, Julia Herzen, and Franz
  Pfeiffer.
\newblock Spectral differential phase contrast x-ray radiography.
\newblock {\em IEEE transactions on medical imaging}, 2019.

\bibitem{kay1993fundamentals}
Steven~M Kay.
\newblock Fundamentals of statistical signal processing, volume i: Estimation
  theory (v. 1).
\newblock {\em PTR Prentice-Hall, Englewood Cliffs}, 1993.

\bibitem{schlomka2008experimental}
JP~Schlomka, E~Roessl, R~Dorscheid, S~Dill, G~Martens, T~Istel, C~B{\"a}umer,
  C~Herrmann, R~Steadman, G~Zeitler, et~al.
\newblock Experimental feasibility of multi-energy photon-counting k-edge
  imaging in pre-clinical computed tomography.
\newblock {\em Physics in Medicine \& Biology}, 53(15):4031, 2008.

\bibitem{ehn2017basis}
S~Ehn, T~Sellerer, K~Mechlem, A~Fehringer, M~Epple, J~Herzen, F~Pfeiffer, and
  PB~No{\"e}l.
\newblock Basis material decomposition in spectral ct using a semi-empirical,
  polychromatic adaption of the beer--lambert model.
\newblock {\em Phys Med Biol}, 62:N1--N17, 2017.

\bibitem{chen2011scaling}
Guang-Hong Chen, Joseph Zambelli, Ke~Li, Nicholas Bevins, and Zhihua Qi.
\newblock Scaling law for noise variance and spatial resolution in differential
  phase contrast computed tomography.
\newblock {\em Medical physics}, 38(2):584--588, 2011.

\bibitem{chabior2011signal}
Michael Chabior, Tilman Donath, Christian David, Manfred Schuster, Christian
  Schroer, and Franz Pfeiffer.
\newblock Signal-to-noise ratio in x ray dark-field imaging using a grating
  interferometer.
\newblock {\em Journal of applied physics}, 110(5):053105, 2011.

\bibitem{revol2010noise}
Vincent Revol, Christian Kottler, Rolf Kaufmann, Ulrich Straumann, and Claus
  Urban.
\newblock Noise analysis of grating-based x-ray differential phase contrast
  imaging.
\newblock {\em Review of Scientific Instruments}, 81(7):073709, 2010.

\bibitem{yashiro2008efficiency}
Wataru Yashiro, Yoshihiro Takeda, and Atsushi Momose.
\newblock Efficiency of capturing a phase image using cone-beam x-ray talbot
  interferometry.
\newblock {\em JOSA A}, 25(8):2025--2039, 2008.

\bibitem{ge2014cramer}
Yongshuai Ge, Ke~Li, and Guang-Hong Chen.
\newblock Cram{\'e}r-rao lower bound in differential phase contrast imaging and
  its application in the optimization of data acquisition systems.
\newblock In {\em Medical Imaging 2014: Physics of Medical Imaging}, volume
  9033, page 90330F. International Society for Optics and Photonics, 2014.

\bibitem{marschner2016helical}
M~Marschner, M~Willner, G~Potdevin, A~Fehringer, PB~No{\"e}l, F~Pfeiffer, and
  J~Herzen.
\newblock Helical x-ray phase-contrast computed tomography without phase
  stepping.
\newblock {\em Scientific reports}, 6:23953, 2016.

\bibitem{kaeppler2017improved}
Sebastian Kaeppler, Jens Rieger, Georg Pelzer, Florian Horn, Thilo Michel,
  Andreas Maier, Gisela Anton, and Christian Riess.
\newblock Improved reconstruction of phase-stepping data for talbot--lau x-ray
  imaging.
\newblock {\em Journal of Medical Imaging}, 4(3):034005, 2017.

\bibitem{rajbhandary2017effect}
Paurakh~L Rajbhandary, Scott~S Hsieh, and Norbert~J Pelc.
\newblock Effect of spatio-energy correlation in pcd due to charge sharing,
  scatter, and secondary photons.
\newblock In {\em Medical Imaging 2017: Physics of Medical Imaging}, volume
  10132, page 101320V. International Society for Optics and Photonics, 2017.

\bibitem{wang2011pulse}
Adam~S Wang, Daniel Harrison, Vladimir Lobastov, and J~Eric Tkaczyk.
\newblock Pulse pileup statistics for energy discriminating photon counting
  x-ray detectors.
\newblock {\em Medical physics}, 38(7):4265--4275, 2011.

\bibitem{brendel2016penalized}
Bernhard Brendel, Maximilian Teuffenbach, Peter~B No{\"e}l, Franz Pfeiffer, and
  Thomas Koehler.
\newblock Penalized maximum likelihood reconstruction for x-ray differential
  phase-contrast tomography.
\newblock {\em Medical physics}, 43(1):188--194, 2016.

\bibitem{donath2009inverse}
Tilman Donath, Michael Chabior, Franz Pfeiffer, Oliver Bunk, Elena Reznikova,
  Juergen Mohr, Eckhard Hempel, Stefan Popescu, Martin Hoheisel, Manfred
  Schuster, et~al.
\newblock Inverse geometry for grating-based x-ray phase-contrast imaging.
\newblock {\em Journal of Applied Physics}, 106(5):054703, 2009.

\bibitem{chabior2011beam}
Michael Chabior, Tilman Donath, Christian David, Oliver Bunk, Manfred Schuster,
  Christian Schroer, and Franz Pfeiffer.
\newblock Beam hardening effects in grating-based x-ray phase-contrast imaging.
\newblock {\em Medical physics}, 38(3):1189--1195, 2011.

\bibitem{scharf1991statistical}
Louis~L Scharf.
\newblock {\em Statistical signal processing: detection, estimation, and time
  series analysis}, volume~63.
\newblock Addison-Wesley, 1991.

\bibitem{roessl2009cramer}
E~Roessl and C~Herrmann.
\newblock Cram{\'e}r--rao lower bound of basis image noise in multiple-energy
  x-ray imaging.
\newblock {\em Physics in Medicine \& Biology}, 54(5):1307, 2009.

\bibitem{weber2011noise}
Thomas Weber, Peter Bartl, Florian Bayer, J{\"u}rgen Durst, Wilhelm Haas, Thilo
  Michel, Andre Ritter, and Gisela Anton.
\newblock Noise in x-ray grating-based phase-contrast imaging.
\newblock {\em Medical physics}, 38(7):4133--4140, 2011.

\bibitem{thuering2014performance}
T~Thuering and M~Stampanoni.
\newblock Performance and optimization of x-ray grating interferometry.
\newblock {\em Philosophical Transactions of the Royal Society A: Mathematical,
  Physical and Engineering Sciences}, 372(2010):20130027, 2014.

\bibitem{engelhardt2008fractional}
M~Engelhardt, C~Kottler, O~Bunk, C~David, C~Schroer, Joachim Baumann,
  M~Schuster, and F~Pfeiffer.
\newblock The fractional talbot effect in differential x-ray phase-contrast
  imaging for extended and polychromatic x-ray sources.
\newblock {\em Journal of microscopy}, 232(1):145--157, 2008.

\bibitem{wang2010analysis}
Zhili Wang, Peiping Zhu, Wanxia Huang, Qingxi Yuan, Xiaosong Liu, Kai Zhang,
  Youli Hong, Huitao Zhang, Xin Ge, Kun Gao, et~al.
\newblock Analysis of polychromaticity effects in x-ray talbot interferometer.
\newblock {\em Analytical and bioanalytical chemistry}, 397(6):2137--2141,
  2010.

\bibitem{paganin2006coherent}
David Paganin.
\newblock {\em Coherent X-ray optics}.
\newblock Number~6. Oxford University Press on Demand, 2006.

\bibitem{boone1992parametrized}
John~M Boone.
\newblock Parametrized x-ray absorption in diagnostic radiology from monte
  carlo calculations: Implications for x-ray detector design.
\newblock {\em Medical physics}, 19(6):1467--1473, 1992.

\bibitem{kottler2010dual}
Christian Kottler, Vincent Revol, Rolf Kaufmann, and Claus Urban.
\newblock Dual energy phase contrast x-ray imaging with talbot-lau
  interferometer.
\newblock {\em Journal of Applied Physics}, 108(11):114906, 2010.

\bibitem{avinash1988principles}
Avinash~C. Kak and Malcolm Slaney.
\newblock {\em Principles of computerized tomographic imaging}.
\newblock IEEE press New York, 1988.

\bibitem{siewerdsen2002framework}
JH~Siewerdsen, IA~Cunningham, and DA~Jaffray.
\newblock A framework for noise-power spectrum analysis of multidimensional
  images.
\newblock {\em Medical physics}, 29(11):2655--2671, 2002.

\end{thebibliography}
\end{document}